\documentstyle[12pt]{article}
\newcommand{\be}{\begin{equation}}   \newcommand{\ee}{\end{equation}}
\newcommand{\bear}{\begin{eqnarray}}
\newcommand{\eear}{\end{eqnarray}}
\newcommand{\ba}{\begin{array}}      \newcommand{\ea}{\end{array}}

\begin{document}
\par \vskip .05in
\begin{titlepage}
\begin{flushright}
FERMI--PUB--00/027--T; 
EFI--2000--3\\
Los Alamos archiver: Physics/0001061\\
 Submitted to {\em The Physics Teacher} \\
{\today}
\end{flushright}
\vfill
\begin{center}
{ 
\Large \bf 
Teaching Symmetry in the Introductory\\
\vskip .1in
Physics Curriculum}
\end{center}
\par \vskip .1in \noindent
\begin{center}
{\bf \large   Christopher T.~Hill$^{1,2}$ \\
\vskip .03in and \\
\vskip .08in Leon M. Lederman$^{1,3}$}
  \par \vskip .1in \noindent
{$^1$Fermi National Accelerator Laboratory\\
 P.O. Box 500, Batavia, Illinois, 60510}
  \par \vskip .05in \noindent
{$^2$The University of Chicago\\
Enrico Fermi Institute, Chicago, Illinois, 60637}
  \par \vskip .05in \noindent
{$^3$ The Illinois Math-Science Academy\\
Aurora, Illinois, USA, 60506-1000 } 
\end{center}     
\par \vskip .05in
\begin{quote}
   {\normalsize
Modern physics is largely defined by fundamental
symmetry principles and N\"{o}ether's Theorem.  
Yet these are not taught, or rarely mentioned, to beginning students,
thus missing an opportunity to reveal that
the subject of physics is as lively and contemporary
as molecular biology, and as beautiful as the arts.  
We prescribe a symmetry module
to insert into the curriculum, of a week's length.
}
\end{quote}
\vfill
\end{titlepage}

\baselineskip=18pt
\pagestyle{plain}
\setcounter{page}{1}

\input math_macros
\def\id{${\bf 1}$}
\def\ra{${\bf R}_{120^\circ}$}
\def\rb{${\bf R}_{240^\circ}$}
\def\ri{${\bf R}_{I}$}
\def\rii{${\bf R}_{II}$}
\def\riii{${\bf R}_{III}$}
\relax
\pagestyle{plain}

\section{Introduction}

Symmetry is a crucial concept in mathematics,
chemistry, and biology.  
Its definition is also applicable to art, music, 
architecture and the innumerable patterns designed by nature, 
in both animate and inanimate forms. 
In modern physics, however,
symmetry may be the most crucial concept of all.  Fundamental
symmetry principles dictate the basic laws of physics,
control stucture of matter, and define
the fundamental forces in nature. 

Some of the most famous mathematicians and physicists had this 
to say about symmetry: 

\begin{itemize}

\item ``I aim at two things: On the one hand to clarify, step by step, the
philosophic-mathematical significance of the idea of symmetry and, on the other,
to display the great variety of applications of symmetry in the arts, in
inorganic and organic nature."  

--- Hermann Weyl [1].

 
\item ``Special relativity emphasizes, in fact is built on, Lorentz symmetry or
Lorentz invariance, which is one of the most crucial concepts in 20th Century
Physics." 

--- C. N. Yang (Nobel Laureate in Physics) [2].

\item ``Symmetry is fascinating to the human mind; everyone likes objects or patterns
that are
in some way symmetrical.... but we are most interested in the symmetries that
exist in the basic laws themselves."

--- Richard P. Feynman (Nobel Laureate in Physics) [3]. 

\item ``I heave the basketball; I know it sails in a 
parabola, exhibiting perfect symmetry, which is
interrupted by the basket.  
Its funny, but it is always interrupted by the basket."

--- Michael Jordan (retired Chicago Bull) [4].
\end{itemize}

Today we understand that all of the fundamental forces
in nature are unified under one elegant symmetry
principle. 
We revere the fundamental symmetries of nature 
and we have come to intimately
appreciate their subtle consequences.  As we
will see, to succumb to
a crack-pot's invention requiring us to give up the law
of energy conservation would be to give up the notion 
of a symmetry principle, that
time flows with no change in the laws
of physics. 
Symmetry controls physics in a most profound way, and this
was the  ultimate lesson of the 20th century.

Yet, even
a sampling of the crucial role of symmetry in physics by beginning
students is completely omitted, not only from the high school
curriculum,
but in the standard first year college calculus-based physics course. 
It does not appear in the Standards.  

It is possible, nonetheless, 
to incorporate some of the underlying ideas of symmetry 
and its relationship to
nature into the beginning courses in physics and mathematics,
at the high school and early college level.  They really are
not that difficult. 
When the elementary courses are spiced with these
ideas, they begin to take on some of the dimensions 
of a humanities or fine arts study: {\em Symmetry is one of
the most beautiful  concepts, 
and its expression in nature is perhaps the most stunning
aspect of our physical world. 
} 

What follows is a description of
a high school module
that introduces the key ideas which, in many examples, ties physics
to astrophysics, biology and chemistry.  
It also reveals 
some of the modern thinking in a conversational way.
We are experimenting in the classroom, 
in Saturday Morning Physics
at Fermilab, and elsewhere in the implementation of this approach.

A lot more material that cannot be presented in this 
brief article, 
can be found at our symmetry website, 
{\bf www.emmynoether.com}.
We will continually 
update our website
as our educational experiment in Symmetry proceeds. 
We encourage you, and your students, to visit it.
And don't hesitate to send us suggestions, comments, and even 
kindly worded complaints.

\section{What is Symmetry?}

When a group of students is asked to define ``symmetry" the
answers they give are generally all correct. 
For example, to the question: ``what
is symmetry?'' we hear some of the following:

\begin{itemize}

\item ``its like when the sides of an equilateral triangle are all the same, or
when the angles are all the same..."

\item ``things are in the same proportion to each other... "

\item ``things that look the same when you see them 
from different points of view ... "

\end{itemize}

From the many diverse ways of describing symmetry, one quickly gets 
to agreement with the scientists' definition: 

\begin{quote}

``Symmetry is an invariance of an object or system to a set of changes
(transformations)."

\end{quote}

In simpler language, a thing
({\em a system}) is said to possess a symmetry if one can make a change 
({\em a transformation}) in the
system such that, after the change, the thing appears exactly the same 
({\em is invariant}) as 
before.   Let us consider some examples.

\subsection{ Translations in Space }

A physical
system can simply be moved from one place to
another place in space.  This is
called a ``spatial translation''.

Consider a classroom pointer.  Usually it is a wooden stick
of a fixed length, about $1$ meter. We can translate the pointer
freely in space. Do its physical properties change as we perform
this translation?  Clearly they do not. The physical material, 
the atoms, the arrangement of atoms into molecules, into the fibrous
material that is wood, etc., do not vary in any obvious
way when we translate the pointer.  This is a symmetry: it
is a statement that the laws of physics themselves are symmetrical
under translations of the system in space. Any {\em equation} we
write describing the quarks, leptons, atoms, molecules, stresses
and bulk moduli, electrical resistance, etc., 
of our pointer must {\em itself}  be invariant under
translation in space.

For example, we can easily write a formula
for the ``length'' of the pointer that is
independent of where the pointer is
located in space (we leave this as an
exercise, or see our website).
Such a formula contains the
information that the length of the pointer, a physical
measure of the pointer,  doesn't change
under translations in space.  Or, put another way,
the formula is ``invariant under spatial
translations.''  While this would be a simple example, the 
(highly nontrivial) assumption is that {\em all correct
equations in physics are translationally invariant!}
Thus, if we have a physics laboratory in which all kinds
of experiments are carried out and all sorts of laws of nature are
discovered and tested, the symmetry dictates that
the same laws will be true if the laboratory itself is
moved ({\em translated }) to another location in space.

The implications of this ``oh so simply stated'' symmetry
are profound.  It is a statement about the nature
of space. 
If space had at very short distances the structure of,
e.g., a crystal, 
then moving from 
a lattice site to a void would change the laws of nature 
within the crystal.  
The hypothesis that space is 
{\em translationally invariant} is equivalent 
to the statement that one point in space is equivalent 
to any other point, i.e. the symmetry is such that 
translations of any system or, equivalently, 
the translation of the coordinate system,  
does not change the laws of nature.  
We emphasize that {\em this is a statement about space itself};
one piece of space is as good as another! We say that space is
smooth or homogeneous (Einstein called it a ``continuum"). 
Equivalently, the laws and the equations that express these laws 
are invariant to translations, i.e., possess translational symmetry.

Now, one can get confused in applying translational invariance.
Consider an experiment to study the translational
symmetry of the electric charge by measuring, e.g., 
acceleration of electrons
in a cathode ray tube. If there were, outside of the laboratory, a huge
magnet, then the experimental results would change when the
tube is moved around inside of the lab. This is not, however,
a violation of translational symmetry, because we forgot to include
the magnet in the move. If we live in a region of space with
intrinsic magnetic fields, then we might detect dependence upon position,
and the symmetry would not seem to exist. However, it is our
belief that flat space is smooth and homogeneous.  
The most profound evidence
comes later.

\subsection{ Translations in Time }

The physical world is actually a fabric of events. 
To describe events we 
typically use a 3-dimensional 
coordinate system for space, but we also need an
extra 1-dimensional
coordinate system for time. This is achieved by building a
clock. The time  on the clock, together with
the 3-dimensional position of something, forms
a four coordinate thing $(x,y,z,t)$, called
an ``event'' (Note: we always assume that the clock is ideally
located at the position of the event, so we don't get confused about
how long it takes for light to propagate from the face
of a distant clock to the location of the event, etc.).
Some examples of events:  (i) We can say that there was 
the event of the firecracker 
explosion at $(x_f, y_f, z_f, t_f)$, (ii) 
The N.Y. Yankees' third baseman hits a fast
pitch at $(x_{H}, y_{H}, z_{H}, t_{H})$, (iii)
Niel Armstrong's foot first touched the surface of the Moon at
the event $(x_M, y_M, z_M, t_M)$.

Now we have the important symmetry hypothesis of physics: 
{\em The laws of physics, and thus all correct
equations in physics,  are invariant under translations
in time}. 
That is, to all of our fabric of events, such as
the events we described above, we can just shift every
time coordinate by an overall common constant.
Mathematically, we replace every time $t_i$ for every
phsyical event by a new
value
$t_i+T$.  The $T$'s cancel in all correct
physics equations; the equations are all {\em time
translationally invariant!} Time, we believe, 
is also smooth and homogeneous.

Indeed, the constancy of the basic
parameters of physics, e.g.,  electric charge, electron
mass, Planck's constant, the speed of light, etc., over vast distances and
times has been established in astronomical and geological
observations to a precision of approaching $10^{-8}$
over the entire age of Universe [2]. The laws of physics
appear to be constant 
in time. The experimental evidence is very strong!

\subsection{ Rotations }

A sphere (or a spherical system, or MJ's basketball) 
can be rotated about any axis that
passes through the center of the sphere.  The rotation angle can be anything
we want, so let's take it to be 63$^o$. After this rotation (often called
an ``operation" or ``transformation") the appearance of the sphere is
not changed. We say that the sphere is ``invariant" under the ``transformation"
of rotating it about the axis by 63$^o$.  Any mathematical description
we use of the sphere will also be unchanged (invariant) under this rotation.
There are an infinite number of
symmetry operations that we can perform upon the sphere. Furthermore,
there is no ``smallest'' nonzero rotation that we can perform; we
can perform ``infinitesimal'' rotations of the sphere.  
We say that the symmetry of the sphere 
is  ``continuous".

Consider again our classroom pointer.  We can rotate the pointer
freely in space. Do its physical properties change as we perform
this rotation?  Clearly they do not. This too is a symmetry: it
is a statement that the laws of physics themselves are symmetrical
under rotations in space.
Under
rotations in free space the length of 
our classroom pointer, $R$, doesn't change.

We could actually perform
a mathematical rotation about the origin
of our coordinate system in which we have written a
formula for the length of a pointer.
We would find that the formula doesn't change (just the coordinates,
the things the formula acts upon, do; this isn't hard to
see, and we do it on the website).
We say that
the length of the pointer is invariant under rotations.
Indeed,  it is our firm belief that 
{\em the laws of physics, and thus all correct
equations in physics,  are invariant under rotations
in space }. 
This is, again, based upon experimental data. It is a statement about
the nature of space; space is said to be {\em isotropic}, that is,
all directions of space are equivalent.

\noindent
In summary:

{\em The  laws of physics  
are invariant under spatial and temporal 
translations, and rotations in space}. 

Needless to say, there are many additional symmetries,
some of which we will discuss later.

\section{Symmetries of the Laws of Physics\\ and 
Emmy N\"{o}ether's Theorem}

In 1905, a mathematician named 
Emmy (Amalie) N\"{o}ether, Fig.(1),
proved the following theorem:

\begin{quote}

{\bf For every continuous symmetry of the laws of physics, 
there must exist a conservation law.}  

{\bf For every conservation law, 
there must exist a continuous symmetry.}
\end{quote}

Thus, we have a deep and profound connection between a symmetry of 
the laws of physics, and the existence of a
corresponding conservation law.
In presenting N\"{o}ether's  theorem at this
level we usually state it without proof
(A fairly simple proof can be given if the student is familiar with
the action principle; it can, however, be motivated with simple
examples, as we do below). 
 
Conservation laws, like the conservation of energy, momentum and 
angular momentum (these are the most famous), are studied in 
high school. They are usually presented as
consequences of Newton's Laws (which is true). 
 We now see from N\"{o}ether's theorem 
that they emerge from symmetry concepts far 
deeper than Newton's laws.  

Now, as we have stated above,
it is an experimental fact that 
the laws of physics are invariannt under 
the symmetry of spatial translations.
This is a strong statement. What is
the physical consequence of this? 
Thus comes the amazing theorem of Emmy N\"{o}ether, 
which states, in this case:

\begin{quote}
 {\bf 
 The conservation law corresponding to space translational symmetry 
is the Law of Conservation of Momentum. }
\end{quote}

So, we learn in senior physics class that the total momentum 
of an isolated system remains constant. The ith element of
the system has a momentum in Newtonian physics of the form:
$ \vec{p}_i = m\vec{v}_i $
and the total momentum
is just the sum of all of the elements,
\beq
\vec{P}_{total} = \vec{p}_1 + \vec{p}_2 + ... + \vec{p}_N
\eeq
for a system of N elements. N\"{o}ether's theorem states
that $\vec{P}_{total}$ is conserved, i.e., it
does not change in time,  no matter how the various
particles interact, because the interactions are determined
by laws that don't depend upon where the whole system is 
located in space!

Note that momentum is, and must be, a vector quantity
(hence the little arrow, $\vec{}\;\;$, over the stuff in the equations).
Why? Because momentum is associated with translations in space,
and the directions you can translate (move) a physical system 
form a vector!  So, if you remember the N\"{o}ether theorem, you
won't forget that momentum is a vector when taking an SAT test!

Turning it around, 
the validity of the Law of Conservation of Momentum
as an observational fact, 
via N\"{o}ether's theorem, supports the hypothesis 
that space is homogeneous, i.e., possessing translational symmetry. 
The more we verify the law of conservation of momentum, 
and it has been tested literally trillions of times in laboratories 
all over the world, at all distance scales,
the more we verify the idea that space 
is homogeneous, and not some kind of crystal lattice!	

We have also stated above the laws of
physics are invariant under translations in time.  
What conservation law then follows by N\"{o}ether's Theorem?  
Surprise!  It is nothing less than the law of conservation of energy:

\begin{quote}
 {\bf 
 The conservation law corresponding to time translational symmetry 
is the Law of Conservation of Energy. }
\end{quote}
  
Since the constancy of the total energy of a system 
is extremely well tested experimentally, this tells 
us that nature's laws are invariant under time translations.

Here is a cute example of how time invariance and energy 
conservation are interrelated.  
Consider a water tower that can hold
a mass  $M$ of water and has a height of 
$H$ meters. 
Assume that the gravitational constant, 
which determines the acceleration of gravity, 
is $g$, on every day of the week, except
Tuesday when it is a smaller value $g'< g$. 
Now, we run water down from the water tower
on Monday through a turbine (a fancy water wheel) generator
which converts the potential energy 
$MgH$ to electrical current to
charge a large storage battery, Fig.(2).
We'll assume $100$\% efficiencies for everything,
because we are physicists. This is Monday's job.
For Tuesday's job we pump the water back up to  $H$,  
using the battery power that we accumulated from Monday's job 
to run the pump.  
But now the  $g'$  value is smaller
than $g$ and  
the work done is $Mg'H$, which is now much less  
than the energy we got from Monday's job. 
This leaves us with $M(g-g')H$ extra energy 
still in the battery, which we
can sell to a local power company to live
on until next Monday.   
This is a perpetual motion machine!  It produces
energy for us, and we can convert that to cash. 
It does not conserve energy because we cooked
up false laws
of physics, in this case gravity, that are not
time translationally invariant! Hence, we violated
a precept of N\"{o}ether's Theorem. (Can you come
up with similar cute example of violating momentum conservation 
by making the laws of physics spatially inhomogeneous?)

We also live in a world where
the laws of physics are rotationally invariant:
\begin{quote}
 {\bf 
 The conservation law corresponding to rotational symmetry 
is the Law of Conservation of Angular Momentum. }
\end{quote}

Conservation of angular momentum is often demonstrated in lecture  
by what is usually called ``the 3 dumbbell experiment".   
The instructor stands on a rotating table, 
his hands outstretched, with a heavy dumbbell in 
each hand (who is the third dumbbell?), Fig.(3).  
He turns slowly, and then brings his hands (and dumbbells) 
close to his body, Fig.(4).  His rotation speed (angular velocity) 
speeds up substantially.  What is kept constant is 
the angular momentum, $J$,
the product of 
$I$, the moment of inertia, times the angular velocity $\omega$.  
By bringing his dumbbells in close to his body, $I$  
is decreased.  
But  $J$,  the angular momentum, must be conserved,
so $\omega$ must increase.   Skaters do this 
trick all the time.   

Atoms, elementary
particles, etc.,  all have angular 
momentum. The intrinsic angular momentum
of an elementray particle is called spin.
In any reaction or collision, the final angular momentum 
must be equal to the initial angular momentum.  
Like our planet earth, particles spin and execute orbits 
and both motions have associated angular momentum.
Data over the past 70 or so years confirms conservation 
of this quantity on the macroscopic scale of people 
and their machines and on the microscopic scale of particles.  
And now, (thanks to Emmy) we learn that these data 
imply that space is isotropic;  
All directions in space are equivalent.

The translational and rotational symmetries of
space and time need not have existed. 
That they do is the way nature is.  
These are some of the actual properties of 
the basic
concepts we use to describe the world: space and time.  

\section{Beyond}

We have described how the fundamental  
conservation laws of everyday physics
follow from the continuous
symmetry properties of space and time.
There are, however, many other conservation laws
that are not usually studied in a first year physics course.
A simple example is the conservation of electric charge
in all reactions. The total electric charge in
an isolated system is a constant in time.
For example, processes like:
\beq
\makebox{electron}^-  \rightarrow  \makebox{neutrino}^0 + \makebox{photon}^0
\eeq
(where superscripts denote charges)
in which case electric charge could completely
disappear, are forbidden. 
On the other hand,
processes like this one do occur:
\beq
\makebox{electron}^- + \makebox{proton}^+ \rightarrow  
 \makebox{neutron}^0  + \makebox{neutrino}^0
\eeq
Since the final state is electrically neutral,
the negative electric charge of the electron must
be identically equal and opposite to 
that of the proton to an infinite number of 
significant figures. Indeed, we can place
a large quantity of Hydrogen gas into a container and
observe to a very high precision that Hydrogen atoms
(which are just bound states of $e^- + p^+$)
are electrically neutral.

This conservation law, by N\"{o}ether's Theorem,
also arises from a profound symmetry
of nature called ``{\em gauge} symmetry.'' 
This is an example of an abstract
symmetry that does not involve space and time. 
In the late 20th century we
have come to realize that all of the forces in nature are controlled by  
such gauge symmetries.  Gauge symmetry is very special, and
it actually leads us to
the complete theory of electrons and photons,
known as (quantum) electrodynamics, which
has been tested to $10^{-12}$ precision. This is the
most accurate and precise theory of nature that 
humans have ever constructed. Perhaps the most
stunning result of the 
20th Century has been the understanding
that all known forces in nature are described by gauge symmetries.

Einstein's Special Theory of
Relativity is all about relative motion, and is based upon
a fundamental symmetry principle
about motion itself. This is a statement that the laws
of physics  must be the same for all observers independent of
their state of uniform motion.
This symmetry  
principle of Relativity can be expressed in a
way that shows that it is a generalization of the concept
of a rotation.  The time interval between two events that occur at the
same point in space is called the ``proper time.''
The proper time can be expressed, like the length of our pointer,
in such a way that it is invariant under motion\footnote{A moving observer 
sees the two events at different points in space; the formula
for proper time involves the spatial separation of the events
divided by $c$, the speed of light; this differs from Newtonian
physics in which the proper time would be independent of the spatial
separation of events.} 
A formula can be written that
relates the coordinate systems of
two observers moving
relative to one another, in terms of their
relative velocity $v$. 
The formula is called a ``Lorentz Transformation''
and it
mixes time
and space, much like a rotation in
the $xy$ plane mixes $x$ and $y$.
Like a rotation, it leaves the proper time invariant.
Thus motion is sort-of like a 
rotation in space and time! 
Unfortunately, we must send you off to a textbook
(or our website)
on Special Relativity to learn about all of the miraculous effects
that occur as a consequence of this.  
Relativity is an expansion of our understanding of the
deep and profound symmetries of nature. 

Other extremely important symmetries arise at the
quantum level.  A simple example is the replacement
of one atom, say a Hydrogen atom sitting in a molecule, by
another Hydrogen atom.  This is a symmetry because all Hydrogen atoms
are exactly the same, or {\em identical}, in all respects. 
There are no warts or moles or identifying body markings on Hydrogen atoms,
or any other atomic scale particle for that matter, such as electrons, protons, quarks, etc. 
The effects of the symmetry associated with
exchanging positions or motions of identical atoms 
or electrons or quarks, has profound effects upon the structure
of matter, from the internal structure of a nucleus of an atom, 
to the properties of everyday materials like metals, insulators
and semiconductors, to the external stucture of a neutron star 
or white dwarf star.
This identical particle symmetry explains
nothing less than the ``Periodic Table of the Elements,'' i.e., how 
the motion and distributions of the electrons 
are organized within the atoms as we go from Hydrogen to
Uranium!  All of chemistry is controlled by the interactions
of electromagnetism together with the symmetry of identical particles. 

In a one week long module we can develop these and other
important symmetries, such as mirror symmetry, time reversal
symmetry and the symmetry between matter and anti-matter
(we can even explain why antimatter exists, from the symmetry of
Relativity and N\"{o}ether's theorem!). 
Conservation laws that are associated with more
abstract symmetries, such as
``quark color,''  and ``supersymmetry,''
can also be illustrated and discussed. Indeed, this leads us
to the frontier of theoretical
physics, e.g., superstrings, M-theory, and the deeply
disturbing open questions, such as ``(why) is the
Cosmological Constant zero?''
As we progressively proceed to the deepest foundations of
the structure of matter, energy, space and time, we must
be more
descriptive for our beginning
students, but we become more thrilled,
enchanted and excited by the fundamental symmetries
that control the structure and evolution of our Universe.

Let this provocative 
conclusion close this very incomplete survey of the role 
of symmetry in physics.  Please visit {\bf
www.emmynoether.com} for a much expanded version of this 
brief letter.

\vspace*{1.0cm}
\noindent
{ \large \bf Acknowledgements }

\noindent
We wish to thank Shea Ferrel for  Figures (3) and (4).  

\newpage

\frenchspacing
\vspace*{1cm}
\noindent
{\Large \bf Bibiliography}
\vspace*{0.5cm}
\begin{enumerate}

\item 
{\em Symmetry}, Hermann Weyl  (Princeton Science Library, 
Reprint edition, Princeton, 1989).

\item C. N. Yang, in {\em Proceedings of the First Int'l Symposium
on Symmetries in Subatomic Physics}, ed. W-Y. Pauchy Huang,
Leonard Kisslingler, v. 32, no. 6-11
(Dec. 1994) 143

\item R. P. Feynman, {\em The Feynman Lectures on Physics},
(Addison-Wesley Pub. Co., 1963).

\item M. Jordan, {\em private communication}.

\item For limits on time dependence of fundamental
constants see,
e.g., F.W. Dyson, in: {\em Aspects of Quantum Theory},
eds. A. Salam and E.P. Wigner (Cambridge Univ. Press,
Cambridge, 1972)
213; in: {\em Current Trends in the the Theory of Fields}
eds. J.E. Lannutti and P.K. Williams (American Institute
of Physics, New York, 1978)
163. See also,
C. T. Hill, P. J. Steinhardt, M. S. Turner 
{\em Phys.Lett.} {\bf B252},1990, 343, and references
therein.

\item see, e.g. {\em Women in Mathematics}, L.M.~Osen,
        {\sl MIT Press} (1974) 141.

\end{enumerate}
\newpage

\begin{figure}[t]
\vspace{6cm}
\includegraphics{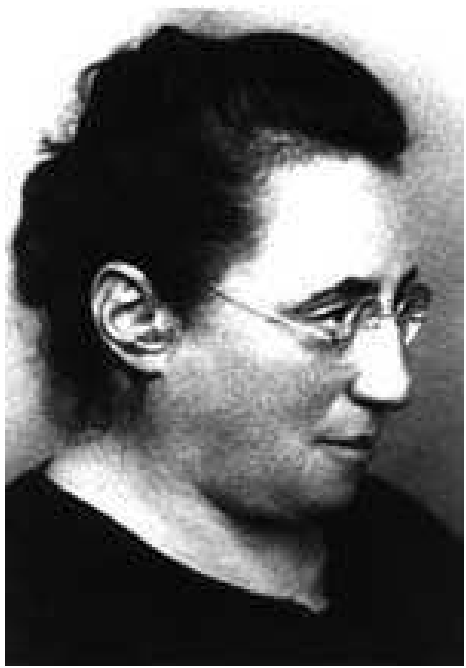}
\vspace{3cm}
\caption[]{ Emmy Noether, 
pronounced
like ``mother.''  Born, 1882,
she practiced at G\"{o}ttingen where
the great mathematicians Hilbert and Klein and the
physicists Heisenberg and Schr\"{o}edinger were professors.
Fleeing the
rise of Naziism, she spent her last 
few years in the U.S. at Bryn Mawr and the Institute
for Advanced Study at Princeton. She
died in 1935 [6]. Emmy N\"{o}ether was one of the greatest
mathematicians of the 20th century.} 
\end{figure}

\begin{figure}[t]
\vspace{12cm}
\includegraphics{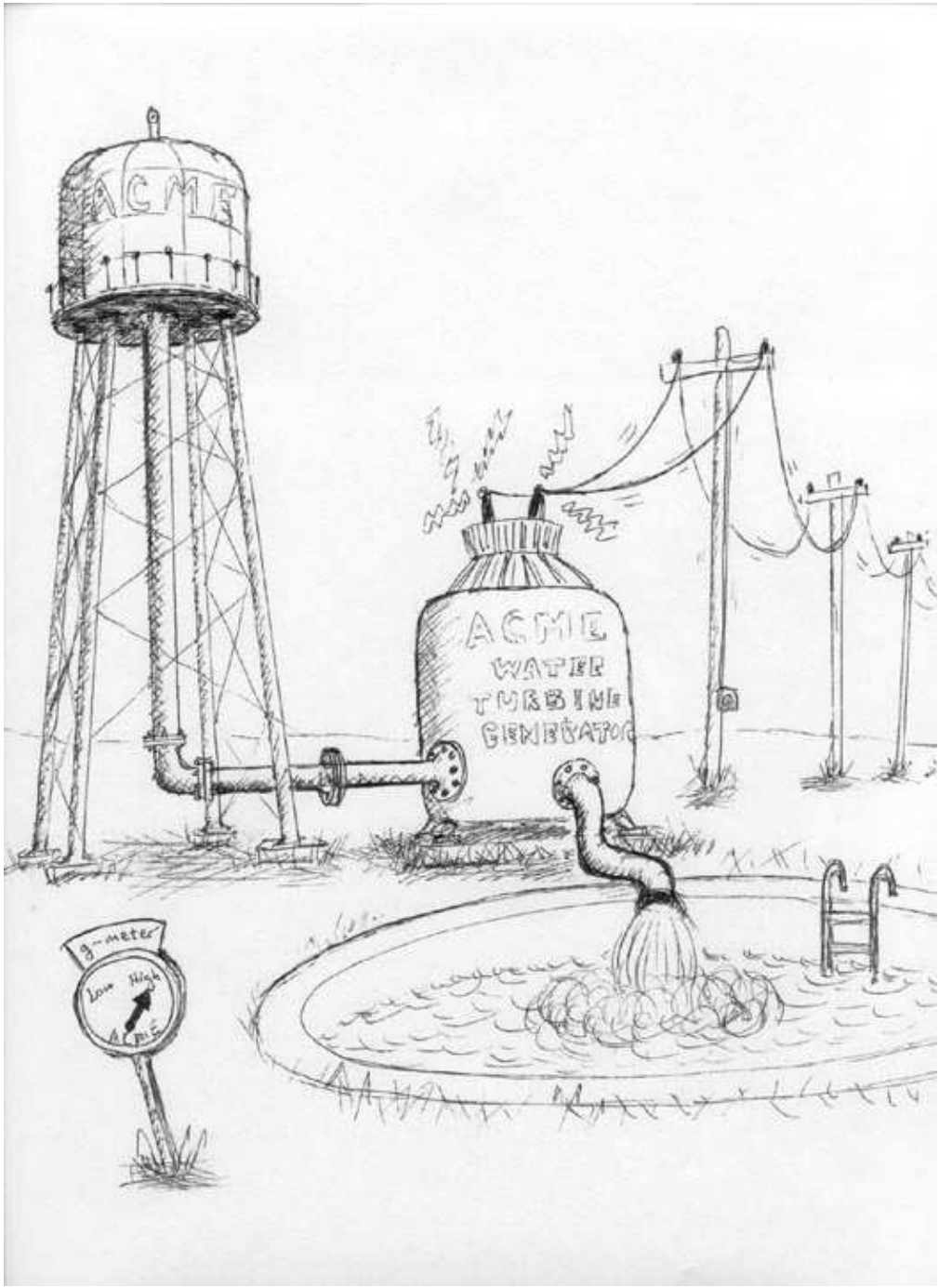}
\vspace{2cm}
\caption[]{  
Water is drained through turbine generator
on days when the gravitational acceleration
is $g' > g$ and energy is produced and sold to power company. On
days when gravitational acceleration is $g< g'$ the water is pumped back
up into the tower at a reduced cost in energy.  
Hence net energy is available
from the system if $g$ is time dependent. } 
\end{figure}

\begin{figure}[t]
\vspace{7.5cm}
\includegraphics{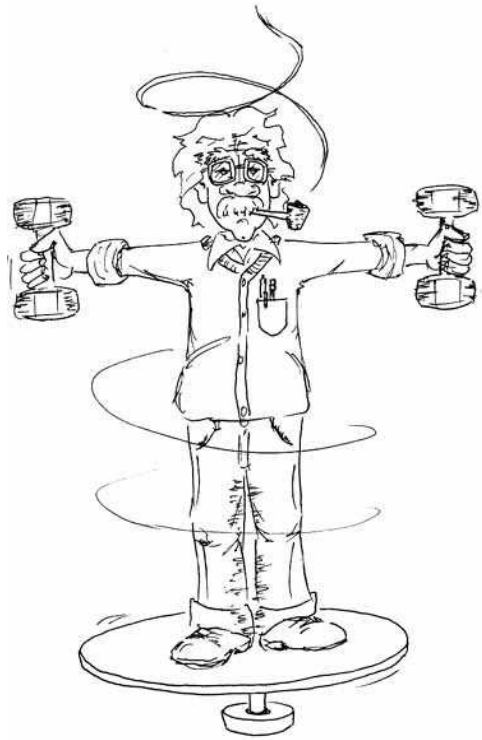}
\vspace{1cm}
\caption[]{ 
The Professor with dumbells rotates slowly when his arms are
outstretched. } 
\end{figure}

\begin{figure}[t]
\vspace{7.5cm}
\includegraphics{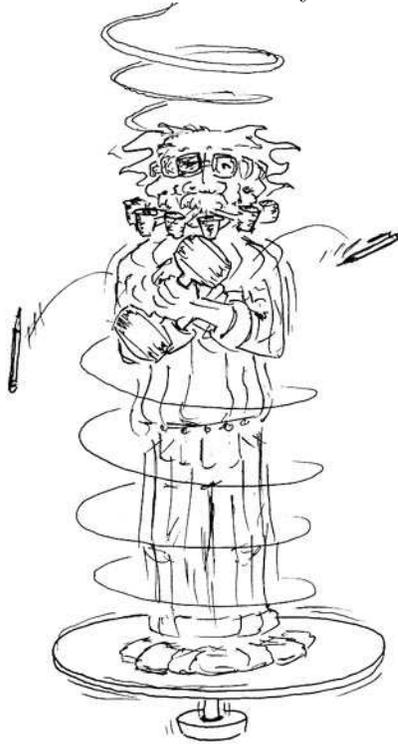}
\vspace{1.5cm}
\caption[]{ 
Pulling the dumbells close to his body reduces the
moment of inertia, but angular momentum is conserved, hence
the Professor rotates faster.  } 
\end{figure}

\end{document}